\documentclass[12pt]{article}
\usepackage{amssymb}
\usepackage{amsmath}
\usepackage{graphicx}
\def\beq{\begin{equation}}
\def\eeq{\end{equation}}
\usepackage[all,2cell,dvips]{xy}

\def\IR{\relax{\rm I\kern -.18em R}}

\begin{document}
\title{On the hamiltonian formulation of an octonionic integrable extension for the Korteweg-de Vries equation}

\author{ {\Large M. Fern\'andez$^{1}$, A. Restuccia$^{2}$ and A. Sotomayor$^{1}$}}
\maketitle{\centerline {$^1$Departamento de Matem\'{a}ticas,
		Universidad de Antofagasta, Chile }}
\maketitle{\centerline{$^2$Departamento de F\'{\i}sica,
		Universidad de Antofagasta, Chile}}
\maketitle{\centerline{e-mail: mercedes.fernandez@uantof.cl, alvaro.restuccia@uantof.cl,
		adrian.sotomayor@uantof.cl }}

\begin{abstract}We present in this work the hamiltonian formulation of an octonionic extension for the Korteweg-de Vries equation. The formulation takes into account the non commmutativity and non associativity of the implicit algebra which defines the equation. We also analize the Poisson structure of the hamiltonian formulation. We propose a parametric master Lagrangian which contains the two hamiltonian structures of the integrable octonionic equation. 

\end{abstract}


Pacs: 02.30.lk, 11.30.-j, 02.30. Jr, 02.10.Hh, 11.10 Ef

\section{Introduction}
Several extensions of the Korteweg-de Vries (KdV) equation have been proposed since its introduction in \cite{Miura1,Miura2,Miura3,Miura4,Miura5}. Also many interesting developments on integrable systems have been obtained since then. 
Besides, an all time challenge in theoretical physics has been to understand the role of the octonion algebra, the unique nonassociative and noncommutative real normalized division algebra in the modelling of the fundamental interactions in nature. In particular, it is well known the relation of the octonions with the maximal supersymmetric Yang-Mills theory in ten dimensions \cite{Evans} and the Supermembrane theory in eleven dimensions in the context of theories searching for the unification of all fundamental interactions. 

Recently, an extension of the KdV equation with fields valued on a general Cayley-Dickson algebra, in particular the octonion algebra, was proposed \cite{Adrian1}. The B\"{a}cklund  transformation, Gardner parametric system, Lax pair and an infinite sequence of conserved quantities were obtained. In this paper we analyse the Hamiltonian structure of that  extension for the fields valued on the octonion algebra.

\section{Korteweg-de Vries equation valued on the algebra of
	octonions}We denote by  $u=u(x,t)$ a function valued on the octonionic algebra $\mathbb{O}$. 

If we denote  by $e_i $ (i=1,\ldots,7),  the imaginary basis of the
octonions, $u$ can be expressed by \beq
u(x,t)=b(x,t)+\overrightarrow{B}(x,t), \eeq where $b(x,t)$ 
and $\overrightarrow{B}=\sum_{i=1}^7B_i(x,t)e_i$ are the corresponding
real and imaginary parts of the octonion.

The Korteweg-de Vries (KdV) equation formulated on the algebra of
octonions, with a one non trivial term can be given by
\beq u_t+u_{xxx}+\frac{1}{2}{(u^2)}_x+[v,u]=0; \eeq when
$\overrightarrow{B}=\overrightarrow{0}$ it reduces to the scalar
KdV equation. The field $v$ is an octonionic constant which can be interpreted as a external field.  

In \cite{Adrian1} equation (2) was formulated in the context of a Cayley-Dickson algebra as an integrable system (in the sense of having an infinite sequence of polynomial conserved quantities).

In terms of $b$ and $\overrightarrow{B}$ the
equation can be re-expressed as
\begin{eqnarray}&& b_t+b_{xxx}+bb_x-\sum_{i=1}^7B_iB_{ix}=0,\\ &&
{(B_i)}_t+{(B_i)}_{xxx}+{(bB_i)}_x+\sum_{j,k=1}^7A_jB_kC_{jki}=0.
\end{eqnarray}

Equation (2) is invariant under Galileo transformations. In
fact, if
\begin{eqnarray*}&&\widetilde{x}=x+ct,\\&&\widetilde{t}=t,\\&&\widetilde{u}=u+c, \\&& \widetilde{v}=v,
\end{eqnarray*} where $c$ is a real constant, then $\frac{\partial}{\partial
x}=\frac{\partial}{\partial \widetilde{x}}$ and
$\frac{\partial}{\partial t}=c\frac{\partial}{\partial
	\widetilde{x}}+\frac{\partial}{\partial \widetilde{t}}.$

Additionally, equation (2) is invariant under the subgroup of the group of automorphisms of the octonions $G_2$ which leaves $v$ invariant. 

If under an automorphism we have
\[ u\rightarrow \phi(u)\] then
\[u_1u_2\rightarrow \phi(u_1u_2)= \phi(u_1)\phi(u_2),\]
and consequently, \[
\left[\phi(u)\right]_t+\left[\phi(u)\right]_{xxx}+\frac{1}{2}{\left({\left[\phi(u)\right]}^2\right)}_x
+[v,\phi(u)]=0,\] if $\phi(v)=v.$

The subgroup of $G_2$ with that property is $SU(3)$. Hence, the KdV extension (2) is invariant under the group $SU(3).$

\section{The master Lagrangian for the octonionic KdV equation}From now on we do not include the bracket term $[v,u]$, that is, we consider $v=0$. We will consider the complete extension elsewhere.

We use now the Hemholtz procedure to obtain a Lagrangian density for the generalized Gardner equation  defined by
\beq  r_t+r_{xxx}+\frac{1}{2} \left(rr_x+r_xr\right)-\frac{1}{12}\left(   \left(r^2\right)r_x+r_x\left(r^2\right)\right){\varepsilon}^2=0.    \label{gardner}\eeq
(5) and (2) are related through
\beq u=r+\varepsilon r_x-\frac{1}{6}{\varepsilon}^2r^2,  \label{trgardner}\eeq in the case $v=0$. Thanks to (6), any solution of (5) gives a solution of (2).

From it and following the construction in \cite{Adrian2} we obtain the two Lagrangians associated to the octonionic KdV equation. The master Lagrangian formulated in terms of the Gardner prepotential $s(x,t)$ ,
\[r(x,t)=s_x(x,t),\] is
\[L_\epsilon(s)=\int_{t_i}^{t_f}dt\int_{-\infty}^{+\infty}\mathcal{L}_\epsilon(s)dx,\] where the Lagrangian density is given by \[\mathcal{L}_\epsilon(s)=\mathbb{R}e\left[-\frac{1}{2}s_xs_t-\frac{1}{6}{\left(s_x\right)}^3+
\frac{1}{2}{\left(s_{xx}\right)}^2+\frac{1}{72}\epsilon^2{\left(s_x\right)}^4\right].\]
Independent variations with respect to $s$ yields the generalized Gardner equation \cite{Adrian1}, as expected since it is guaranteed by the Helmholtz procedure.

The Lagrangian density $ \mathcal{L}_\epsilon(s)$ is invariant under the action of the exceptional Lie group $G_2$. In fact, consider the infinitesimal $G_2$ transformation
\begin{eqnarray*}&&s\rightarrow s+\delta s\\&&\delta s=\lambda^{ij}D_{e_i,e_j}(s),\end{eqnarray*} where $\lambda^{ij}$ are the real infinitesimal parameters  and $D_{e_i,e_j}(s)$ are the generators of the associated derivation algebra, the Lie algebra of the Lie group $G_2$. Its definition and properties are discussed, for example in \cite{Adrian1}.

We get
\begin{eqnarray*}&&\delta \mathcal{L}_\epsilon(s)=\mathbb{R}e\left[-\frac{1}{2}{\left(
		\delta s\right)}_xs_t-\frac{1}{2}s_x{\left(
		\delta s\right)}_t-\frac{1}{6}\left({\left(
		\delta s\right)}_x{(s_x)}^2+s_x{\left(
		\delta s\right)}_xs_x+{(s_x)}^2{\left(
		\delta s\right)}_x\right)\right]+\\&&+\mathbb{R}e\left[\frac{1}{2}
	\left({\left(
		\delta s\right)}_{xx}s_{xx}+s_{xx}{\left(
		\delta s\right)}_{xx}s_{xx}\right)+\frac{1}{72}\epsilon^2\left({\left(
		\delta s\right)}_x{(s_x)}^3+s_x{\left(
		\delta s\right)}_x{(s_x)}^2+{(s_x)}^2{\left(
		\delta s\right)}_xs_x+{(s_x)}^3{\left(
		\delta s\right)}_x\right)\right],\end{eqnarray*}
we also have ${\left(
	\delta s\right)}_x=\lambda^{ij}D_{e_i,e_j}(s_x)=\delta s_x$.

Using now the Leibnitz rule for $D_{e_i,e_j}$ we obtain
\begin{eqnarray*}&&\delta \mathcal{L}_\epsilon(s)=\mathbb{R}e\left\{\lambda^{ij}D_{e_i,e_j}
	\left(-\frac{1}{2}s_xs_t-\frac{1}{6}{(s_x)}^3+{(s_{xx})}^2+\frac{1}{72}\epsilon^2{(s_x)}^4\right)\right\}\end{eqnarray*} but
$D_{e_i,e_j}(\hspace{2mm})$ is pure imaginary for any argument, hence $\delta \mathcal{L}_\epsilon(s)=0$.

If we take the limit $\epsilon\rightarrow0$, we obtain a fisrt Lagrangian for the octonionic KdV equation,
\[L(w)=\int_{t_i}^{t_f}dt\int_{-\infty}^{+\infty}dx\mathbb{R}e\left[-\frac{1}{2}w_xw_t
-\frac{1}{6}{(w_x)}^3+\frac{1}{2}{(w_{xx})}^2\right].\]
Independent variations with respect to $w$ yields, using $u=w_x$, the octonionic KdV equation (2).

If we take the following redefinition
\begin{eqnarray*}&&s\rightarrow \hat{s}=\epsilon s\\&&\mathcal{L}_\epsilon(s)\rightarrow\epsilon^2\mathcal{L}_\epsilon(\hat{s})\end{eqnarray*} and take the limit $\epsilon\rightarrow\infty$ we obtain
\[\lim_{\epsilon\rightarrow\infty}\epsilon^2\mathcal{L}_\epsilon(\hat{s})=\mathcal{L}^M(\hat{s}), \] where
\[\mathcal{L}^M(\hat{s})=\mathbb{R}e\left[-\frac{1}{2}{\hat{s}}_x{\hat{s}}_t
+\frac{1}{2}{({\hat{s}}_{xx})}^2+\frac{1}{72}{({\hat{s}}_{x})}^4\right].\] We get in this limit the generalized Miura Lagrangian
\[L^M( \hat{s})=\int_{t_i}^{t_f}dt\int_{-\infty}^{+\infty}dx\mathcal{L}^M(\hat{s}).\]
The Miura equation is then obtained by taking variations with respect to $\hat{s}$, we get
\[\hat{r}_t+\hat{r}_{xxx}-\frac{1}{18}{(\hat{r})}^3_x=0,\hspace{2mm}\hat{r}\equiv\hat{s}_x,\] while the Miura transformation arises after the redefinition process, it is $u=\hat{r}_x-\frac{1}{6}\hat{r}^2.$

Any solution of the Miura equation, through the Miura transformation, yields a solution of the octonionic KdV equation. Since $L_\epsilon(s)$ is invariant under $G_2$, the same occurs for $L(w)$ and $L^M( \hat{s})$ and consequently for the equations arising from variations of them.

The Lagrangian formulation of the octonionic KdV equation may be used as the starting step to obtain the hamiltonian structures of the octonionic system.

We may now construct, using de Dirac approach \cite{Dirac} for constrained systems,  the two hamiltonian structures associated to the lagrangians $L$ and $L^M$. This approach was used in
\cite{Nutku,Kentwell} to obtain the, previously known, first and
second hamiltonian structures of KdV equation. Applications of Dirac's procedure to obtain the corresponding Poisson structures  in another  extensions of KdV type can be seen in \cite{Adrian2,Adrian3}.

We first consider the Lagrangian $L$ and define the conjugate momenta associated to $w_0$ and $w_i (i=1,\ldots,7)$ defined by
\begin{eqnarray*}&&w_0=\mathbb{R}e(w)\\&&w_ie_i=\mathbb{I}m(w), \hspace{2mm}w_i,i=1,\ldots,7;\end{eqnarray*}
summation in the repeated index is understood.

We denote $p$ and $p_i$ the conjugate momenta of $w_0$ and $w_i$ respectively.

We have, from the definition of the momenta,
\begin{eqnarray*}&&p=\frac{\partial \mathcal{L}}{\partial w_{0t}}=-\frac{1}{2}w_{0x}\\&&
	p_i=\frac{\partial \mathcal{L}}{\partial w_{it}}=\frac{1}{2}w_{ix}.\end{eqnarray*}

They define constraints on the phase space
\begin{eqnarray*}&&\phi_0=p+\frac{1}{2}w_{0x}\\&&\phi_i=p_i-\frac{1}{2}w_{ix}, \hspace{2mm}w_i,i=1,\ldots,7.\end{eqnarray*}
These are second class constraints. In fact,
\beq \begin{array}{lll}\{\phi(x),\phi( \hat{x})\}_{PB}=\partial_x\delta(x-\hat{x}),\\
	\{\phi(x),\phi_i( \hat{x})\}_{PB}=0,\\\{\phi_i(x),\phi_j( \hat{x})\}_{PB}=-\delta_{ij}\delta(x-\hat{x}).\end{array}\eeq
In order to obtain the Poisson structure on the constrained phase space we use the Dirac theory of constraints. The Dirac bracket between two functionals $F$ and $G$ is
\beq\{F,G\}_{DB}=\{F,G\}_{PB}-{\left\langle\left\langle\{F,\phi_m(x^\prime\}_{PB}C_{mn}(x^\prime,x^{ \prime\prime})
	\{\phi_n(x^{\prime\prime}),G\}_{PB}\right\rangle\right\rangle}_{x^\prime,x^{ \prime\prime}}\eeq
where ${\{\}}_{x^\prime}$ denotes integration on $ x^\prime$ from $-\infty$ to $+\infty$. The indices $m,n$ range from $0$ to $7$ and $C_{mn}$ are the components of the inverse of the matrix whose components are the poisson brackets of the second class constraints (7).

The Hamiltonian is obtained from the Lagrangian $L$ through a Legendre transformation,
\beq\begin{array}{lll}H=\int_{-\infty}^{+\infty}\mathcal{H}dx,\\
	\mathcal{H}=\mathbb{R}e   \left[\frac{1}{6}{(w_x)}^2-\frac{1}{2}{(w_{xx})}^2\right]=
	\frac{1}{6}{(w_{0x})}^3-\frac{1}{2}{(w_{0xx})}^2-\frac{1}{2}w_{0x}{(w_{ix})}^2+\frac{1}{2}{(w_{ixx})}^2=\\
	=\frac{1}{6}{(u_0)}^3-\frac{1}{2}{(u_{0x})}^2-\frac{1}{2}u_0{(u_i)}^2+\frac{1}{2}{(u_{ix})}^2,\end{array}
\eeq where $u_0$ is the real part of the octonionic field $u$ and $u_i$ the components of its imaginary part. Summation on repeated indices is understood. The Poisson structure on the constrained phase space is obtained from the Dirac brackets. We obtain
\beq \begin{array}{lll}\{u_0(x),u_0( \hat{x})\}_{DB}=-\partial_x\delta(x-\hat{x}),\\
	\{u_i(x),u_j( \hat{x})\}_{DB}=\delta_{ij}\partial_x\delta(x-\hat{x}),\\\{u_0(x),u_i( \hat{x})\}_{DB}=0.\end{array}\eeq
(9) and (10) define the first Hamiltonian structure of the octonionic KdV equation.

We may proceed in an analogous way to construct the second Hamiltonian structure. We start from the Miura Lagrangian $L^M$, define the conjugate momenta, determine the constraints and construct the Dirac brackets of any two functionals. In particular for $u_0$ and $u_i$. We obtain
\beq \begin{array}{lll}\{u_0(x),u_0( \hat{x})\}_{DB}=-\partial_{xxx}\delta(x-\hat{x})+
	\frac{1}{2}u_{0x}\delta(x-\hat{x})+\frac{2}{3}u_0\partial_x\delta(x-\hat{x}),\\
	\{u_i(x),u_j( \hat{x})\}_{DB}=\delta_{ij}\left[-\partial_{xxx}\delta(x-\hat{x})-\frac{1}{3}
	u_{0x}\delta(x-\hat{x})-\frac{2}{3}u_0\partial_x\delta(x-\hat{x})\right],
	\\\{u_0(x),u_i( \hat{x})\}_{DB}=\frac{1}{3}u_{ix}(x)\delta(x-\hat{x})+\frac{2}{3}u_i(x)\partial_x\delta(x-\hat{x}).\end{array}\eeq
The Hamiltonian is $H^M=\int_{-\infty}^{+\infty}\mathcal{H}^Mdx,$ where \beq \mathcal{H}^M=-u_0^2+u_i^2.\eeq
(11) and (12) define the second Hamiltonian structure of the octonionic KdV equation.

\section{Conclusions}We presented an octonionic extension of the Korteweg-de Vries equation with a non trivial term and we considered some of its invariances. Starting with a Lagrangian master formulation, in the case in which the non trivial term is zero (and so we can suppose that $v=0$), we obtained the hamiltonian structures and the consequent Poisson brackets of the system, using the Dirac's method in the analysis of the constraints. We believe that the procedure followed in this work can be used in several of the known KdV coupled systems and in particular we will consider its application elsewhere in the case in which the non trivial term is non zero and also when the underliyng structure is a Cayley-Dickson algebra. 

$\bigskip$

\textbf{Acknowledgments}

M. F., A. R. and A. S. are partially supported by Project Fondecyt
1161192, Chile.

\end{document}